\documentclass[prd,superscriptaddress,nofootinbib,showpacs,preprint]{revtex4}
\usepackage{epstopdf}
\usepackage{graphicx}
\usepackage[usenames,dvipsnames]{color}
\usepackage[normalem]{ulem}
\usepackage{amsmath,amssymb}
\usepackage[colorlinks]{hyperref}
\usepackage{color}

\begin{document}
	
\title {Viscous Self Interacting Dark Matter and Cosmic Acceleration}
\author{Abhishek Atreya }
\email{abhishek.atreya@ipr.res.in}
\affiliation{Basic Theory and Simulation Group, Institute of Plasma 
	Research, Gandhinagar, 382428, India}
\author{Jitesh R Bhatt }
\email{jeet@prl.res.in}
\affiliation{Theory Division, Physical Research Laboratory,
	Navarangpura, Ahmedabad - 380 009, India}	
\author{Arvind Mishra }
\email{arvind@prl.res.in}
\affiliation{Theory Division, Physical Research Laboratory,
	Navarangpura, Ahmedabad - 380 009, India}
\affiliation{Indian Institute of Technology, Gandhinagar, 382424, India }	
	\date{\today}
	 
\def\be{\begin{equation}}
\def\ee{\end{equation}}
\def\bearr{\begin{eqnarray}}
\def\eearr{\end{eqnarray}}
\def\zbf#1{{\bf {#1}}}
\def\bfm#1{\mbox{\boldmath $#1$}}
\def\hf{\frac{1}{2}}
\def\sl{\hspace{-0.15cm}/}
\def\omit#1{_{\!\rlap{$\scriptscriptstyle \backslash$}
		{\scriptscriptstyle #1}}}
\def\vec#1{\mathchoice
	{\mbox{\boldmath $#1$}}
	{\mbox{\boldmath $#1$}}
	{\mbox{\boldmath $\scriptstyle #1$}}
	{\mbox{\boldmath $\scriptscriptstyle #1$}}
}

\begin{abstract}
Self interacting dark matter (SIDM) provides us with a 
consistent solution to certain astrophysical observations in conflict 
with collision-less cold DM paradigm. In this work we estimate the shear 
viscosity $(\eta)$ and bulk viscosity $(\zeta)$
of SIDM, within kinetic theory formalism, for galactic and 
cluster size SIDM halos. To that extent we make use of the recent constraints on SIDM crossections for the dwarf galaxies, LSB galaxies 
and clusters. We also estimate the change in solution of Einstein's 
equation due to these viscous effects and find that 
$\sigma/m$ constraints on SIDM from astrophysical data provide us with 
sufficient viscosity to account for the 
observed cosmic acceleration at present epoch, without the need of any 
additional dark energy component. Using the estimates of dark matter 
density for galactic and cluster size halo we find that the mean free path of 
dark matter $\sim$ few Mpc. Thus the smallest scale at which the viscous 
effect start playing the role is cluster scale. Astrophysical data for dwarf, 
LSB galaxies and clusters also seems to suggest the same. The entire 
analysis is independent of any specific particle physics motivated 
model for SIDM.

 \end{abstract}
  
  
 \maketitle

 \section{Introduction} 

  The self interacting dark matter (SIDM) is a lucrative alternative to the
standard collision-less cold dark matter (CDM) paradigm. The motivation for
SIDM comes from the small scale observations of the Universe which are
inconsistent with the CDM predictions. There are four major problems viz,
core-cusp problem, diversity problem, missing satellites problem and
too-big-to-fail problem. SIDM paradigm has the potential to provide a unifying
solution to these problems. We refer to \cite{Tulin:2017ara} for a detailed
review on SIDM and its possible resolution to the above mention problems.

An important implication of self interactions in the DM sector would be a
non-zero equation of state. It has been argued that one should attribute
thermodynamic properties like internal energy with DM owing to its
self-interactions \cite{Kleidis:2011ga}. In addition, DM can also be provided
with a polytropic EoS \cite{Kleidis:2014xia}. Both the above attributes are
capable of explaining the Type Ia supernovae (SNe-Ia) data without any
additional dark energy component in cosmic fluid. The condition of negative
pressure is naturally satisfied in these models. There are indications already
that dark sector may have a negative equation of state (EoS) at cluster scale
\cite{MNL2:MNL21082} and at cosmological scale \cite{Calabrese:2009zza}.

Another standard approach to obtain a dark energy like feature is to assign a
non-zero bulk viscosity to the dark sector
\cite{Murphy:1973zz,Padmanabhan:1987dg,Fabris:2005ts}. For an homogeneous,
isotropic fluid with bulk viscosity $\zeta$, the pressure is
$p_{\mathrm{v}} = p+\Pi_{b}$, where $p_{\mathrm{v}}$ is the pressure of viscous
fluid and $\Pi_{b}=-3 \zeta H$. For sufficiently large $\zeta$, the pressure
would be negative thus mimicking dark energy.
The source of bulk viscosity has been attributed to
neutrinos \cite{Das:2008mj}, exotic scalar fields \cite{Gagnon:2011id} or to
the decay of cold dark matter into relativistic particles \cite{Mathews:2008hk}.
However these models are severely constrained by observations of the large
scale structure formation \cite{Li:2009mf,Piattella:2011bs,Velten:2011bg}.
The reason being that a large bulk viscosity leads to decay of the
gravitational potential during structure formation \cite{Li:2009mf}. 
  
A SIDM particle has the potential to avoid these constraints on the
viscosity as it behaves like a standard CDM candidate at large scales. One can
understand this by looking at the scattering rate of SIDM,
\begin{equation}
R_{\mathrm{scat}} = \frac{\langle\sigma v\rangle\rho_{\mathrm{SIDM}}}{m},
  \label{eq:scrate}
\end{equation}
where $m$ is the mass of SIDM particle, $\langle\sigma v\rangle$ is the
velocity weighted crossection of SIDM particle, $\rho_{\mathrm{SIDM}}$ is the
density of the dark matter. As the density falls, the scattering rate goes to
zero and dark-matter approaches the standard collision-less CDM behaviour,
however at small scales i.e. near the central region of dark matter halo,
the density is large and consequently the scattering rate is higher. It is
thus expected that these scattering processes would lead to non-zero 
viscosity within the dark matter halo.

In this work we estimate the shear and bulk viscosity of dark matter within the
kinetic theory framework. Assuming the velocity distribution of SIDM to be
Maxwellian we obtain the expressions for the shear and the bulk viscosity of
SIDM in terms of its velocity weighted crossection to mass ratio,
$\langle\sigma v\rangle/m$, and the average velocity, $\langle v\rangle$, of
the dark matter halo.
The estimates of $\langle\sigma v\rangle/m$ were obtained in
\cite{Kaplinghat:2015aga} by utilizing the astrophysical data from dwarf
galaxies, LSB galaxies and clusters. Assuming local thermalization in the 
dark matter halo, we use these estimates to infer the shear and bulk 
viscosity in the galaxy as well as in the cluster size dark matter halo. 
We find that the viscous coefficients increase by two orders of magnitude 
from galactic to cluster scales.

As an application of these results we explore the scenario discussed in ref.
\cite{Floerchinger:2014jsa}. There it was argued that the local
velocity perturbations in cosmological fluid depend on the shear and the bulk
viscosity of the dark sector. If the back-reaction of these velocity
perturbations is large enough then they contribute substantially to the
energy dissipation. This modifies the solution to Einstein's equations and 
one could then explain the accelerated cosmic expansion without any dark
energy component. We test this assertion using our results for the shear and
bulk viscosity of the SIDM. We estimate the contribution of the SIDM
viscosity to the energy dissipation and find that the dissipative effects are
sufficient to account for the observed cosmic acceleration.

The organization of the paper is as follows. In section \ref{sec:kin} we will
discuss the calculation of bulk and shear viscosity of the dark matter using
kinetic theory and present our estimates of the shear and bulk viscosity at
the galactic and cluster scales. In section \ref{sec:csmcexp} we estimate the
energy dissipation in viscous hydrodynamics and use Einstein equations to argue
that the dissipative effects due to viscosity can lead to cosmic acceleration. 
We present our results on scattering crossection of SIDM and cosmic acceleration in section \ref{sec:disc} and conclude in section 
\ref{sec:conc}.

\section{Shear and Bulk Viscosity From Kinetic Theory}
\label{sec:kin}

In this section we will estimate the shear and bulk viscosity of SIDM
using the kinetic theory formalism. We work in natural units.
Within the hydrodynamics, the stress energy tensor can, in general, be
decomposed in two parts: $T^{\mu\nu}_{\mathrm{ideal}}$ and $T^{\mu\nu}_{\mathrm{diss}}$,
where $T^{\mu\nu}_{\mathrm{diss}}$ is the dissipative part of energy momentum
tensor. In a local Lorentz frame it can be written as
\begin{equation}
  \label{eq:loclemt}
  T^{ij}_{\mathrm{diss}} = -\eta\left(\frac{\partial u^{i}}{\partial x^{j}}
  + \frac{\partial u^{j}}{\partial x^{i}}
  -  \frac{2}{3}\frac{\partial u^{k}}{\partial x^{k}}\delta^{ij}\right)
  - \zeta\frac{\partial u^{k}}{\partial x^{k}}\delta^{ij},  
\end{equation}
where $\mathbf{u}$ is the fluid velocity, $\eta$ and $\zeta$ are the shear
viscous and bulk viscous coefficients
respectively. To estimate viscous coefficients we use the kinetic 
theory formalism and obtain the expression for the shear and
bulk viscosity of SIDM in terms of its distribution function, and velocity
weighted crossection \cite{Gavin:1985ph,Kadam:2015xsa}.

The starting point of kinetic theory is the Boltzmann equation,
\begin{equation}
 \label{eq:be}
  \frac{\partial f_{p}}{\partial t} + v_{p}^{i}\frac{\partial f_{p}}{\partial x^{i}} = I\lbrace f_{p}\rbrace,
\end{equation}
where $\mathbf{v}_{p}$ is single particle velocity, $f_{p}$ is the distribution
function and $I\lbrace f_{p}\rbrace$ is collisional
integral describing the rate of change of $f_{p}$ due to collisions. One can
evaluate this integral under certain approximations and we use the ``relaxation
time approximation''. In this approximation it is assumed that the variation
of distribution function is slow in space and time or equivalently the system
is close to thermal equilibrium. Thus one can locally assign thermodynamic
quantities like temperature and energy to the system. In addition
it is also assumed 
collisions are effective in bringing the system to thermal equilibrium within
a characteristic time, termed as collision time
$(\tau)$ of the system. In the later stages of cosmic evolution, where it is
assumed that the dark matter halos have virialised and a Maxwellian
distribution can be assigned to the dark matter particles, it seems reasonable
that the relaxation time approximation will hold for SIDM. For a collision-less
dark matter the approximation is clearly invalid.

Under these assumptions the collision term can be approximated by the
linearized form 
\begin{equation}
  \label{eq:rta}
  I\lbrace f_{p}\rbrace \simeq - \frac{\delta f_{p}}{\tau},
\end{equation}
where $\delta f_{p} \equiv (f_{p}-f_{p}^{0})$ is the deviation of distribution
function from the equilibrium distribution function $f_{p}^{0}$. In a general
setup $\tau$ can be a function of energy. Eq. (\ref{eq:rta}) in conjunction
with eq. (\ref{eq:be}) gives
\begin{equation}
 \label{eq:delf}
 \delta f_{p} = - \tau \left(\frac{\partial f_{p}^{0}}{\partial t} +
 v_{p}^{i}\frac{\partial f_{p}^{0}}{\partial x^{i}} \right).
\end{equation}  
Since we assume local equilibrium, we can define the average energy density
$(T^{00})$ and momentum density $(T^{0i})$ of the system using the distribution
function. Extending the definition to the $OJ$ component of $T^{\mu\nu}$,
we can define $T^{OJ}$ as 
\begin{equation}
  \label{eq:emt}
  T^{OJ} = \int\frac{d^{3}p}{(2\pi)^{3}}~v^{i}p^{j}f_{p}.
\end{equation}
Using $f_{p} = f_{p}^{0} + \delta f_{p}$, with $\delta f_{p}$ given by eq.
(\ref{eq:delf}), we get the $T^{OJ}_{\mathrm{Dias}}$ as
\begin{equation}
  \label{eq:tij}
  T^{ij}_{\mathrm{diss}} = - \int \frac{d^{3}p}{(2\pi)^{3}}~\tau v^{i}p^{j}\left(
  \frac{\partial f_{p}^{0}}{\partial t} +
 v_{p}^{l}\frac{\partial f_{p}^{0}}{\partial x^{l}} \right).
\end{equation}

Let us now consider the fluid motion along, say, $x$ axis with fluid
velocity $u_{x}(y)$, i.e. $\mathbf{u} = \left(u_{x}(y),0,0\right)$. In this
case eq. (\ref{eq:loclemt}) reduces to
$T^{xy} = -\eta \partial u_{x}/\partial y$. Using the equilibrium distribution
function to be of the form $f^{0}_{p} = \exp (-p^{\mu}u_{\mu}/T)$ in eq.
(\ref{eq:tij}) and comparing with $T^{xy}$ written above we get
\begin{equation}
  \label{eq:visc}
  \eta = \frac{1}{15T}\int \frac{d^{3}p}{(2\pi)^{3}}~\tau~\frac{p^{4}}{E_{p}^{2}}
  ~\frac{\partial f_{p}^{0}}{\partial E_{p}}.
\end{equation}  

For bulk viscosity we have to take the trace of eq. (\ref{eq:loclemt}) and
compare with trace of eq. (\ref{eq:tij}). Using $T^{\mu\nu}_{~,\nu} = 0$ and
some manipulation one can obtain the expression for the bulk viscosity to be
\begin{equation}
  \label{eq:blk}
  \zeta = \frac{1}{T}\int\frac{d^3p}{(2\pi)^3}\tau\left[E_{p}C_{n}^2 -
    \frac{p^2}{3E_{p}}\right]^2 f_{p}^0,
\end{equation}
%
%
where $C_{n} = \frac{\partial P}{\partial \epsilon}\lvert_{n}$ is the speed of sound at constant number density.
Eqn. (\ref{eq:visc}) and (\ref{eq:blk}) are relativistic expression for the
shear bulk viscosity from kinetic theory. To estimate these quantities for
dark matter we need to estimate the relaxation time $\tau$. 
We may approximate relaxation time with average relaxation time $\tilde{\tau}$
(this average is over energy distribution), given by
\begin{equation}
  \label{eq:avt}
\tilde{\tau}^{-1} = n \langle\sigma v\rangle,
\end{equation}
where $n, ~\langle\sigma v\rangle$ represent average number density and
velocity weighted cross section average. Since we are working in the cold
SIDM paradigm we have to estimate $\eta, \zeta$ in the non-relativistic limit.
We thus use the non-relativistic Maxwell-Boltzmann distribution in fluid rest
frame, with eqn. (\ref{eq:avt}), in the non-relativistic limit of eqn.
(\ref{eq:visc}). This gives us
\begin{equation}
    \label{eq:visc1}
    \eta = \left( \frac{m}{\langle\sigma v\rangle}\right) \left( \frac{T}{m}
    \right).
\end{equation}
We now make an assumption that virial theorem leads to equipartition of 
energy, $\frac{1}{2}m\langle v^2 \rangle \ = \frac{3}{2}T$, 
for the dark matter halos, where $m$ is the mass of dark matter and $T$ is the
virialised temperature of dark matter halo. Also for Maxwellian velocity
distribution $ \sqrt{\langle v^2 \rangle} = 1.08 \langle v \rangle $. Eqn
(\ref{eq:visc1}) thus takes the form
\begin{equation}
\label{eq:visc2}
  \eta =   \frac{1.18m\langle v \rangle^2}{3\langle\sigma v\rangle}.
\end{equation}

 A similar exercise for eqn. (\ref{eq:blk}) gives the bulk viscosity as
\begin{equation}
    \label{eq:blk2}
    \zeta =\frac{5.9 \ m\langle v\rangle^2}{9\langle\sigma v\rangle}.
\end{equation}
%

\begin{figure}
	\begin{center}
	\includegraphics[width=0.45\textwidth]{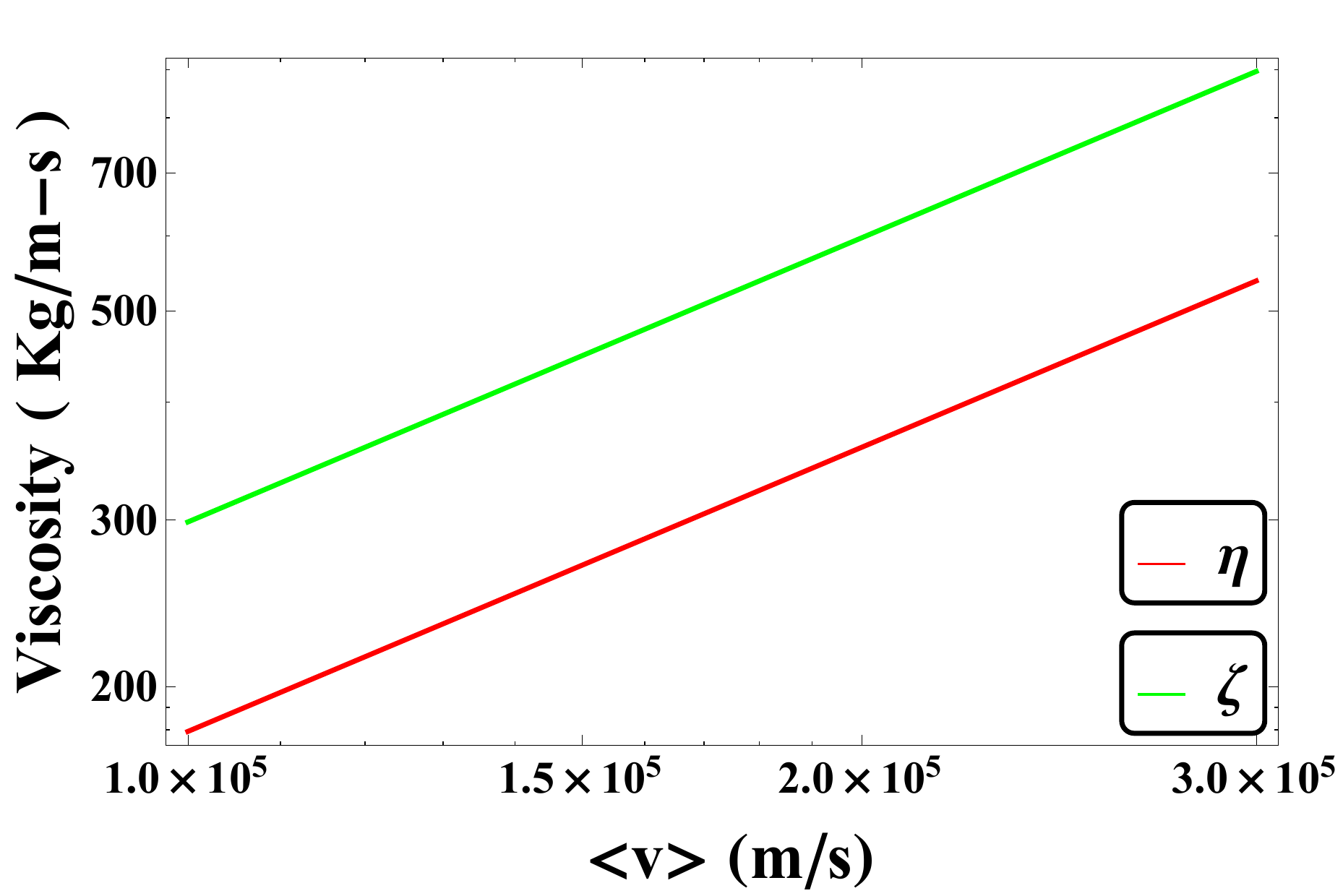}
	\includegraphics[width=0.45\textwidth]{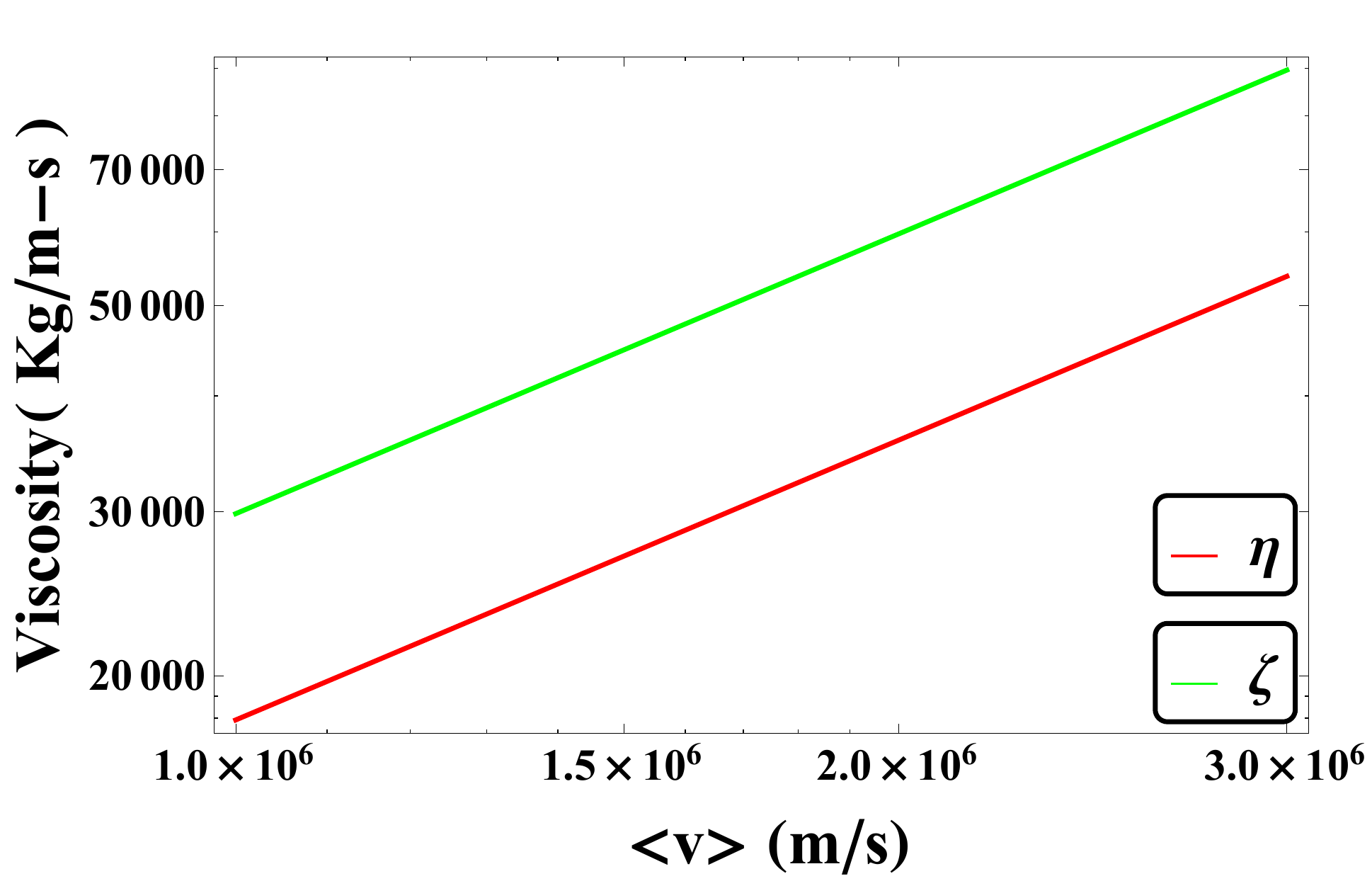}
	\caption{Plots of $\eta$ and $\zeta$ \textit{vs} $\langle v\rangle$
		at galactic (left) and cluster (right) scale.}
		\label{fig:visc}
	\end{center}
 \end{figure}

 We can now estimate the values of shear and bulk viscosity for galactic and 
cluster size dark matter halos. It is clear from equation (\ref{eq:visc2}) and 
(\ref{eq:blk2}) that the viscous coefficient $\eta$ and $\zeta$ depend 
on the mass $m$ of DM candidate, its velocity weighted crossection average 
$\langle\sigma v\rangle$, and on $\langle v \rangle$. Also we note that the 
bulk viscosity contribution is slightly larger than the shear contribution at 
each scale. If we take $\left <\sigma v\right>/(\left< v\right>m)$ as an 
estimation of $\left<\sigma\right>/m$ at a given scale, then $\eta$ and $\zeta$ are proportional to $\left<v\right>$ in a dark matter halo. From galactic to cluster scale 
$\left<\sigma\right>/m$ changes from $\sim 1$ cm$^{2}$g$^{-1}$ to $\sim 0.1$ 
cm$^{2}$g$^{-1}$ \cite{Kaplinghat:2015aga} and $\left< v\right>$ goes from 
$10^{2}$ to $10^{3}$ kms$^{-1}$. This would mean that viscosity ($\eta, \zeta$ 
both) increase by two orders of magnitude as we go from galactic to 
cluster scale, see fig. \ref{fig:visc}. Note that in the plots we have converted the results to SI units while the eqn. (\ref{eq:visc2}) and  
(\ref{eq:blk2}) are derived using natural units.

\section{Effect of Viscosity on Cosmic Expansion}
\label{sec:csmcexp}

In this section we explore the consequence of dark viscosity on the evolution 
of the universe. It was argued in \cite{Floerchinger:2014jsa} that sufficient 
viscous contribution from the dark sector can modify the solution of 
Einstein's equation and may explain the observed accelerated expansion. We use 
the estimates of $\eta$ and $\zeta$, obtained above, to study the extent of 
this affect.

To solve for Einstein's equations we need the energy momentum tensor for a
viscous fluid, which, in Landau frame $(u_{\mu}T^{\mu\nu}=-u^{\nu}\epsilon)$ is
given by the expression
\begin{equation}
   \label{eq:tmunu}
    T^{\mu \nu} = \epsilon u^{\mu} u^{\nu} + (P + \pi_{b}) \Delta^{\mu \nu} +
    \pi^{\mu \nu}.
\end{equation}
Here $ \Delta^{\mu \nu} = u^{\mu} u^{\nu} + g^{\mu \nu}$ is the projection
operator orthogonal to the fluid velocity, $\pi_{b}$ is the bulk stress and
shear stress, $ \pi^{\mu \nu}$, satisfies the conditions
$u_{\mu} \pi^{\mu \nu} = 0, ~\pi_{\mu}^{\mu} = 0$. In the first order
hydrodynamics the bulk and shear stresses are given by the expressions
\begin{subequations}
  \begin{equation}
    \label{eq:pib}
    \pi_{b} = - \zeta \nabla_{\mu} u^{\mu},
  \end{equation}
  \begin{equation}
    \label{eq:pimunu}
    \mathrm{and}~~\pi^{\mu \nu} = - 2\eta\left[ \Delta^{\mu\alpha}\Delta^{\nu\beta} +
      \Delta^{\mu\beta}\Delta^{\nu\alpha} -
      \frac{2}{3}\Delta^{\mu\nu}\Delta^{\alpha\beta}\right] \nabla_{\alpha}u_{\beta}.
  \end{equation}
\end{subequations}

In addition to $G^{\mu\nu} = - 8\pi G T^{\mu\nu}$, we have the covariant
energy momentum conservation $\nabla_{\mu} T^{\mu \nu} = 0$. For the metric we
consider the conformal FRW metric with scalar perturbations. In the limit of
small metric perturbations and small velocities $(v^{2}<<1)$, we get the energy
conservation equation as \cite{Floerchinger:2014jsa}
%
\begin{subequations}
\begin{equation}
 \label{eq:encons}
 \frac{1}{a}\dot{\left<\epsilon\right>}+3H(\left<\epsilon\right>+\left<P\right>
 -3\left<\zeta\right>H) = D,
\end{equation}
\begin{equation}
  \label{eq:diss}
  \mathrm{where}~D = \frac{1}{a^2}\left<\eta\left[\partial_{i}v_{j}
    \partial_{i}v_{j}+\partial_{i}v_{j}\partial_{j}v_{i} -
    \frac{2}{3}\partial_{i}v_{i}\partial_{j}v_{j}  \right]\right>
  + \frac{1}{a^{2}}\left<\zeta[\vec{\nabla}\cdot\vec{v}]^2\right> +
 \frac{1}{a}\left\langle\bar{\vec{v}}\cdot\vec{\nabla}(P-6\zeta H)\right\rangle
\end{equation}
\end{subequations}
is the dissipation term. In the above equations $\left<A\right>$ is the
spatial average of the quantity $A$. As is evident from equation
(\ref{eq:encons}), that the evolution of
average energy density depends crucially on the dissipative term $D$.
Only at late times the velocity gradients acquire non zero values
hence it is at late times that the dissipative term becomes important and 
contribute significantly to the average energy evolution equation. This 
in-turn affects the evolution of Universe through Einstein's equations.

Using the results obtained in the previous sections we would like to come up
with an estimate of $D$ and check the extent of this effect. We use the
following simplifying assumptions: $(i)$ We assume that $\eta,\zeta$ do not
vary over the spatial region which is averaged. $(ii)$ gradients and curls of
velocity fields are prominent at a scale $L$, thus we approximate
$\partial_{i} \sim 1/L\equiv 1/(R_{H}\alpha)$, where $R_{H}=H^{-1}$ is the Hubble
size and $\alpha = L/R_{H}$ is the fraction of Hubble size where the
derivatives are prominent. $(iii)$ We confine ourselves to the present epoch,
thus we set $H = H_{0}$ and scale factor $a\equiv a_{0}=1$.
$(iv)$ The dark equation of state is $P+\pi_{b} = \hat{\omega} \epsilon$. For
cold SIDM, $\hat{\omega}= 0$, thus
$P=-\pi_{b}=\zeta(\nabla\cdot\bar{v} + 3H_{0}) =\zeta
\left[\left(\frac{H_{0}\langle v\rangle}{\alpha}\right) + 3H_{0}\right]$. 
Thus last term in eqn. (\ref{eq:diss}) becomes  
$$\langle\bar{\vec{v}}\cdot\vec{\nabla}(P-6\zeta H)\rangle = \zeta \left( \frac{H_{0}\langle v\rangle}{\alpha}\right)^2\left[ 1 - 3 \ \frac{\alpha}{\langle v\rangle}\right] \sim \zeta \left( \frac{H_{0}\langle v\rangle}{\alpha}\right)^2,$$
since $\alpha/\langle v\rangle \ll 1$. With these assumptions and with the use
of eqn. (\ref{eq:visc2}) and (\ref{eq:blk2}) in the eqn. (\ref{eq:diss}), we
get the value of dissipation term at present epoch as
%
%
%
\begin{equation}
\label{eq:diss2} 
  D  = \frac{16.32\langle v\rangle^4}{9} \left( \frac{m}{\langle\sigma v\rangle}\right)\left(\frac{H_{0}}{\alpha}\right)^2.
\end{equation}
%

To estimate the effect of viscous term on the evolution of the Universe, eqn.
(\ref{eq:encons}), with $D$ given by eqn. (\ref{eq:diss2}), needs to be solved
in conjunction with the suitably averaged Einstein's equation. The simplest
choice is the traced equation
$\left<G^{\mu}_{\mu}\right> = 8\pi G \left<T^{\mu}_{\mu}\right>$, which gives
\begin{equation}
  \label{eq:trace}
  \frac{\ddot{a}}{a^{3}} = \frac{\dot{H}}{a}+2H^{2} = \frac{4\pi G}{3}
  \left(\left<\epsilon\right>-3\left<P\right> -3\left<\pi_{b}\right>\right).
 \end{equation}
%

Using eqn (\ref{eq:encons}), (\ref{eq:diss2}) with  eqn. (\ref{eq:trace})
one can find equation for deceleration parameter (defined as $q = -1 - 
\frac{\dot{H}}{aH^{2}}$) to be
\begin{equation}
- \frac{dq}{d \ln a} + 2 (q-1) \left(q- \frac{1}{2}\right) = \frac{4\pi G D}{3 H^{3}}.
\label{eq:qevol}
\end{equation}
%

From observations, $q\approx-0.6$ at present epoch 
\cite{Ade:2013zuv}. Thus the dissipation term (assuming 
$\left|{\frac{dq}{d\ln a}}\right|\ll 1$ ) is estimated to be 
\cite{Floerchinger:2014jsa}
 $ \frac{4\pi GD}{3H_{0}^{3}} \approx 3.5$.
This implies that the dissipation term $D$ can provide an
explanation for cosmic acceleration if the viscous effects are strong enough
for dark matter. Hence we need to calculate dissipative term $D$, which 
depends on $\langle\sigma v\rangle/m$, $\langle v\rangle$ and the size 
$\left(H_{0}/\alpha\right)$ of SIDM halo. An important consideration here 
is the scale of averaging (galactic or cluster). We know, from observations, 
that the accelerated cosmic expansion manifests itself at super-cluster or 
larger scales, thus intuitively the averaging doesn't seem appropriate for a 
SIDM halo of galactic size. To settle the issue of averaging scales we 
estimate the mean free path of the SIDM. The idea behind this exercise is that 
for the hydrodynamic description to be valid (which allows us to write eqn. 
(\ref{eq:tmunu}) in the first place) the mean free path should be smaller 
than, or at least be the order of, the averaging scale. 

  As the densities for galaxy and clusters are traditionally quoted in the 
  units of M$_{\odot}$kpc$^{-3}$, we leave natural units for estimations of 
  mean free path. From eqn. (\ref{eq:visc2}) we have $D\sim\eta 
  \left(\vec{\nabla}\cdot\vec{v}\right)^{2}$. For $\eta$ we use the textbook 
  expression for viscosity of a dilute gas \textit{viz.} 
  $\eta = \rho v \lambda/3$. With 
  approximations $\partial_{i}\sim 1/L$ and $v \sim \langle v\rangle$ we get
  $D \sim \rho \lambda \langle v\rangle^{3}/3 L^{2}$. Equating with 
  eqn. (\ref{eq:diss2}) we get
  %
 $ \lambda \sim 5 \left(\frac{m}{\sigma}\right)\frac{1}{\rho}.$
%
 Since $\sigma/m$ is expressed in units of cm$^{2}$/g, we convert it to 
kpc$^{2}/$M$_{\odot}$. In these units the expression for mean free path is
\begin{equation}
\lambda \sim 5\times 10^{10}\left(\frac{m}{\sigma}\right)\frac{1}{\rho} ~~ (\textrm{in~kpc}),
  \label{eq:mfp}
\end{equation}
where $\sigma/m$ corresponds to the numerical value in the units of 
cm$^{2}$/g and $\rho$ is numerical value in units of M$_{\odot}$kpc$^{-3}$.
For galactic scale $\sigma/m \sim 2    $ cm$^{2}$/g \cite{Kaplinghat:2015aga}. 
So for the SIDM mean free path of $\sim 10$ kpc would require that the density 
at the galactic scales be $2.5\times10^{^9} $ M$_{\odot}$kpc$^{-3}$. At 
galactic scale the densities are much lower $e.g$ for Milky way the estimates 
for DM density (with NFW profile) are $\sim 10^{6}-10^{7}$ M$_{\odot}$kpc$^{-3}$
\cite{Weber:2009pt} which are much lower. For LSB galaxies the peak 
density for isothermal case are $\sim 10^{7}$ M$_{\odot}$kpc$^{-3}$ 
\cite{KuziodeNaray:2007qi}. For dwarf galaxies the dark matter concentration 
can be a bit larger ($\sim 10^{8}$ M$_{\odot}$kpc$^{-3}$) \cite{Oh:2010ea},
but the typical sizes are smaller than $10$kpc. 

For cluster scale halos, $\sigma/m \sim 0.1$ cm$^{2}$/g 
\cite{Kaplinghat:2015aga}, thus for $\lambda \sim 1$ Mpc we need the halo 
density to be $\sim 5\times 10^{8}$ M$_{\odot}$kpc$^{-3}$ [from eqn (\ref{eq:mfp})]. Such high densities have been estimated in cluster size 
dark matter halos using isothermal
profile for dark matter \cite{Newman:2012nv,Newman:2012nw}. It thus seems 
appropriate to argue that the hydro description and consequently the 
averaging in dissipation term is not appropriate for galactic scale SIDM 
halos but for cluster scale at the least.


%
\begin{figure}[t!]
	\begin{center}
		\includegraphics[width=0.48\textwidth]{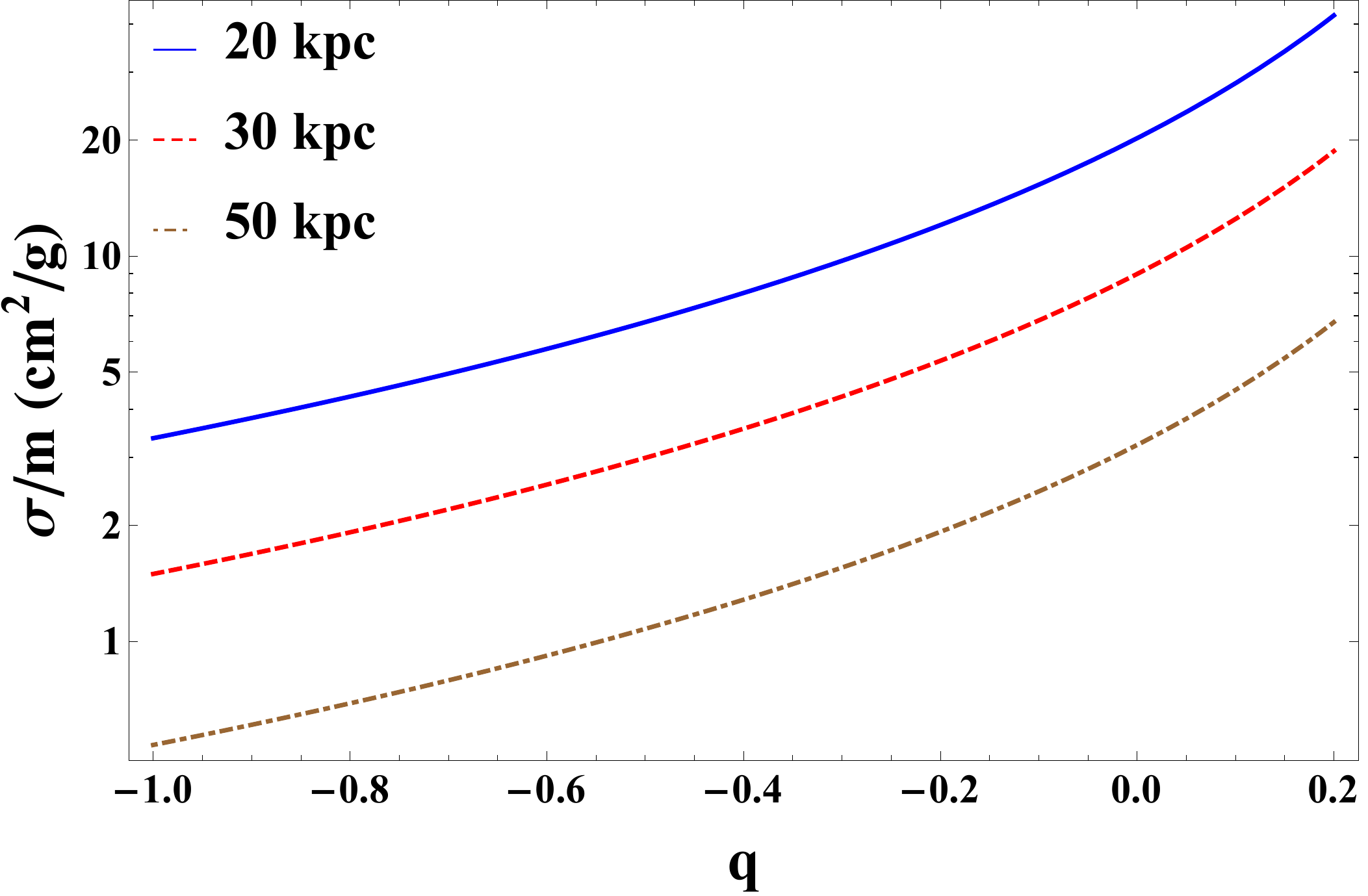}
		\includegraphics[width=0.48\textwidth]{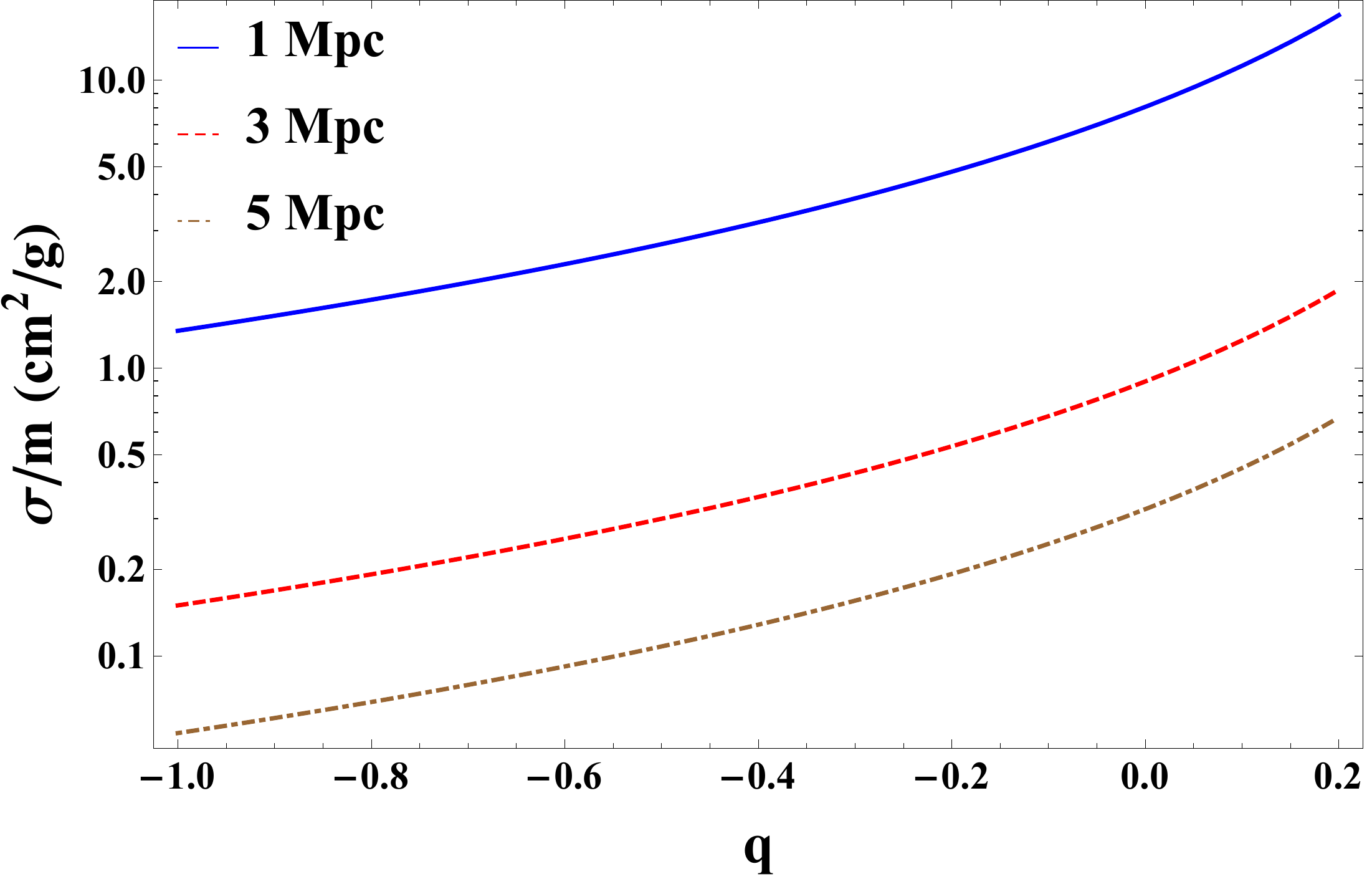}
		\caption { The plot between $\sigma/m$ vs $q$ for different 
			$\langle v\rangle $ and $\alpha$ at galactic (\textit{left}) and cluster
			(\textit{right}) scale. The velocities are taken to be $100$ kms$^{-1}$ 
			and $1000$ kms$^{-1}$ for galactic and cluster scales. Larger negative 
			value of $q$ is supported by smaller values of $\sigma/m$.}
		\label{fig:q}
	\end{center} 
\end{figure}
\section{Results and Discussions}
\label{sec:disc}

\begin{figure}[t!]
	\includegraphics[width=0.45\textwidth]{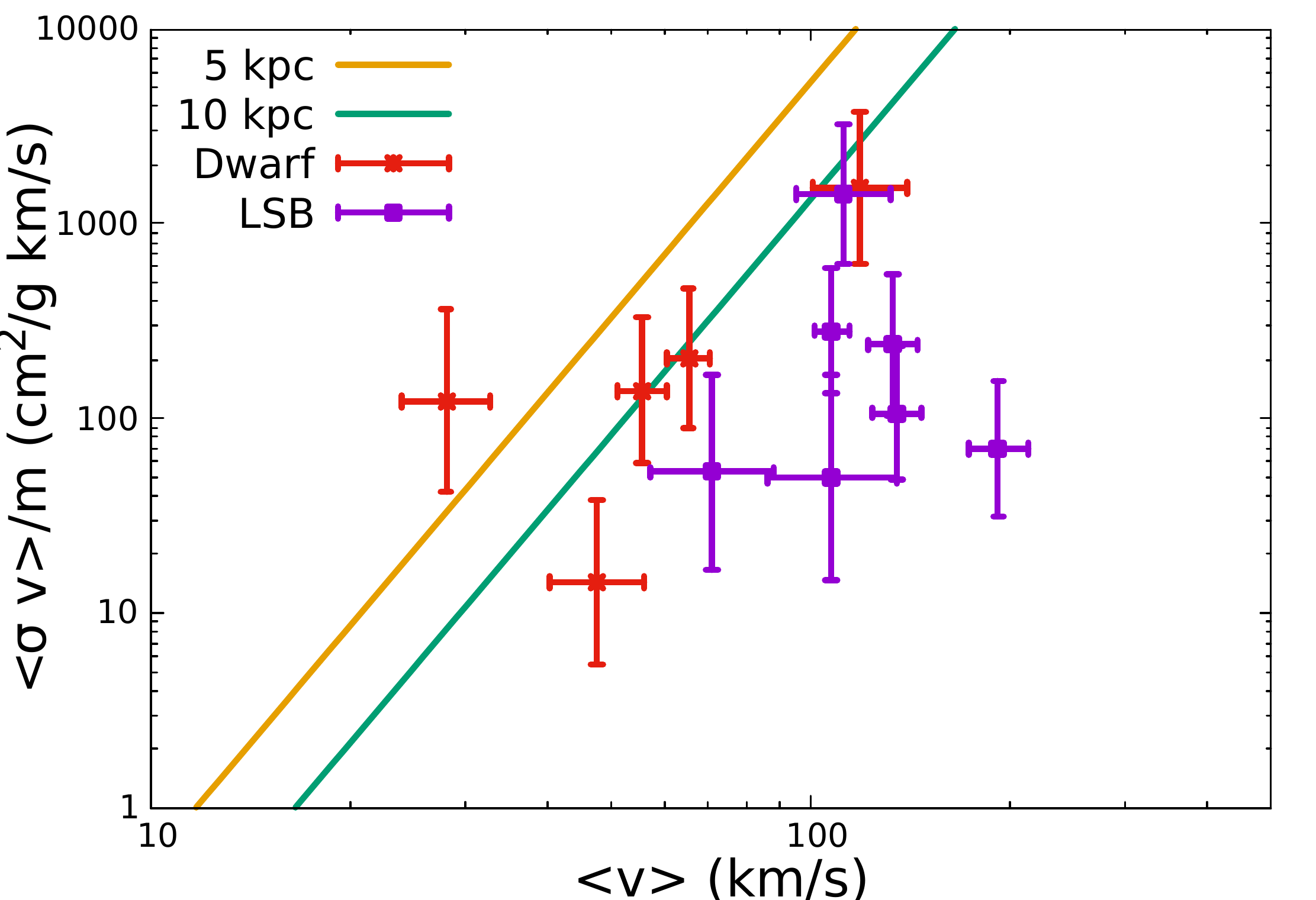}
	\includegraphics[width=0.45\textwidth]{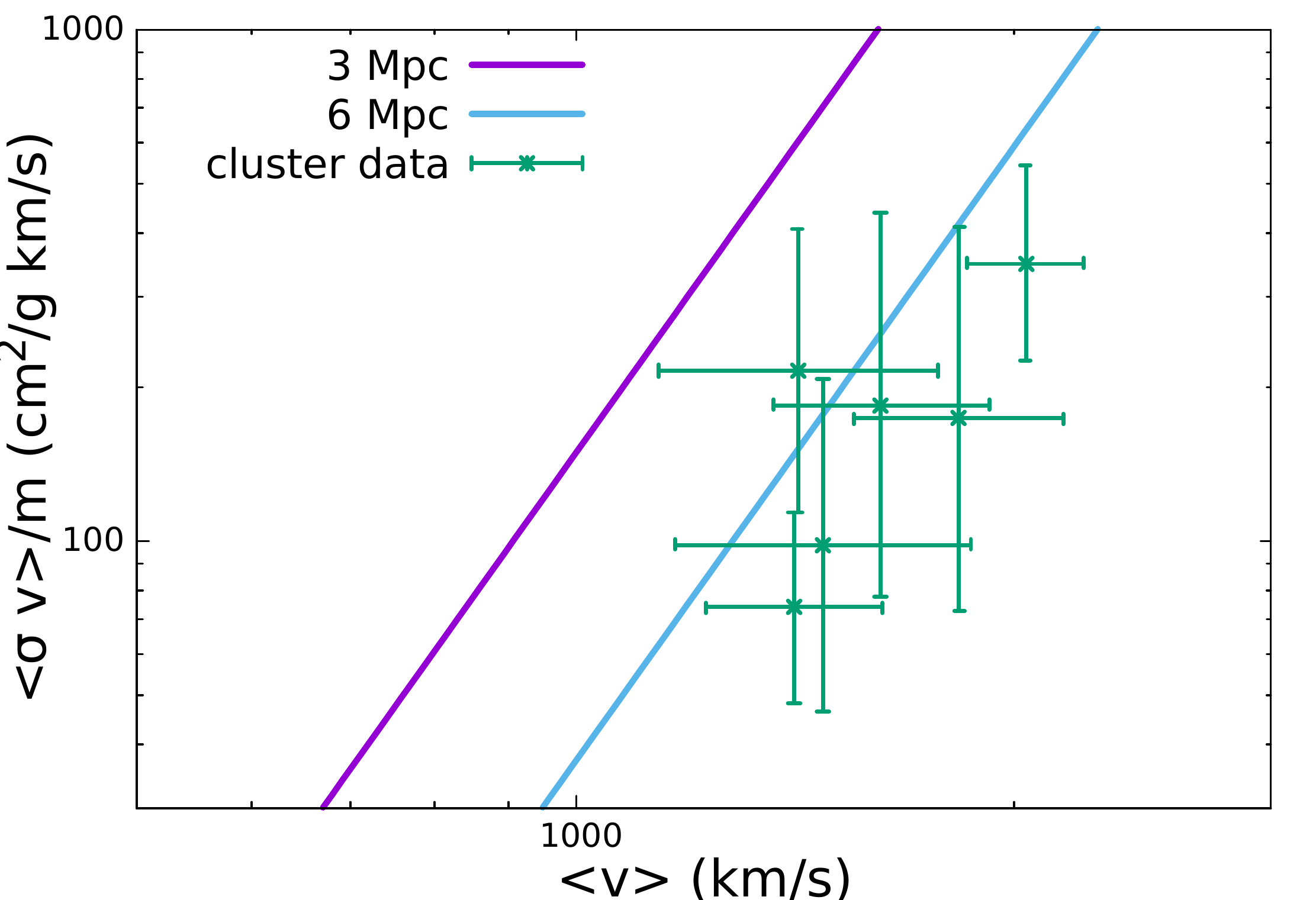}
	\caption{Plot of $\langle\sigma v\rangle/m$ \textit{vs} $\langle v\rangle$ 
		for galactic (\textit{left}) and cluster (\textit{right}) size dark halo 
		for $ q = - 0.6$. The data points are from \cite{Kaplinghat:2015aga}.}
	\label{fig:svbym}
\end{figure}

The evolution of the deceleration parameter $q$ is governed by eqn 
(\ref{eq:qevol}), with $D$ given by eqn (\ref{eq:diss2}). 
Assuming $\left|{\frac{dq}{d\ln a}}\right|\ll 1$, we get 
$\langle\sigma v\rangle /m$ as
\begin{equation}
\frac{\langle\sigma v\rangle }{m} = \frac{65.28\pi \langle v\rangle^4}{27(q-1)(2q - 1)H_{0}m_{pl}^2\alpha^2} ,   \ \ \textrm{where} \  m_{pl}^{2} \equiv \frac{1}{G}.
\label{eq:svbym}
\end{equation}
  Dividing both sides by $\langle v\rangle$ provides 
  us with an estimate of $\sigma/m$ for different halo sizes. The result 
  is presented in fig. \ref{fig:q}. For plotting we choose $\langle v\rangle$
  to be $100$ kms$^{-1}$ for galactic scale while for a cluster size halo 
  we take $\langle v\rangle \sim 10^{3}$ kms$^{-1}$ as a representative 
  number. It is evident that present observed deceleration parameter 
  $(q \simeq -0.6)$ can be obtained for a dark matter halo of roughly 
  $50-30$ kpc size with $\sigma/m\sim 1-2$ cm$^2$/g. Similarly at cluster 
  scale the dark matter halo of $3-5 \ Mpc $ size can provide us with 
  correct $q$ values with $\sigma/m \sim 0.1$ cm$^2$/g. The $\sigma/m$ 
  values of $1-2$ cm$^2$/g at galactic scale and $0.1$ cm$^2$/g at cluster 
  scale were deduced from astrophysical data in \cite{Kaplinghat:2015aga}. 
  To check it further, we plot 
  $\langle\sigma v\rangle/m$ \textit{vs} $\langle v\rangle$ 
  for $q = - 0.6$, at galactic and cluster scales, and 
  compare with the $\langle\sigma v\rangle/m$ values inferred from the 
  astrophysical data \cite{Kaplinghat:2015aga} in fig. \ref{fig:svbym}.

%
  The large dispersion in $\langle\sigma v\rangle/m$ at the galactic scale, 
  as compared to the cluster scale, is an important feature of fig. 
  \ref{fig:svbym}. The size of dwarf galaxies is almost $5$ kpc and the 
  LSBs are also of size smaller than $10$ kpc. Eqn (\ref{eq:svbym}) thus 
  provides a poor fit to the data as can be seen in fig \ref{fig:svbym}. 
  However for cluster scale halos the scatter is comparatively less and the 
  fit is better than the galactic scale. This can be thought of as the better 
  applicability of the averaging process at the cluster scale as compared to 
  the galactic scale as discussed at the end of previous section.

\section{Conclusions}
 \label{sec:conc}

  In this work we have estimated the shear and bulk viscosity due to dark 
  matter self interactions within kinetic theory formalism. Assuming 
  local thermalization in dark matter halos, we determined the relation 
  between viscous coefficients $(\eta,\zeta)$, velocity 
  weighted crossection to mass ratio $(\langle\sigma v\rangle/m)$ of 
  dark matter and $\langle v\rangle$ of dark matter halos. 
  The estimates suggest that 
  $\eta$ and $\zeta$ change by roughly two orders of magnitude from the 
  galactic to cluster scale. We also looked at the effect of dissipation 
  due to these viscous effects, on cosmic acceleration. We find that 
  $\eta$ and $\zeta$ estimated from astrophysical data for $\sigma/m$ 
  of SIDM might account for the observed cosmic acceleration.
  To get a better understanding of the scales where these effects might be 
  important, we also estimated the mean free path of SIDM. We find 
  that the mean free path $\sim $few Mpc for a cluster size dark matter halo 
  is supported by the astrophysical 
  data. We thus conclude that smallest scale for viscous effects to play 
  a role in dynamics of the Universe should be cluster scale. We set our 
  theoretical understanding against the astrophysical data and find that 
  data also points in the same direction.


\section{Acknowledgement}
 We would like to thank Subhendra Mohanty and Namit Mahajan for useful 
 discussions and comments. We are also thankful to Manoj Kaplinghat for providing the data of $\langle\sigma v\rangle/m$ and $\langle v\rangle$ from 
 their simulations. AM would also like to thank Sampurnanand, Bhavesh 
 Chauhan and Richa Arya for fruitful discussions. A major part of this work was carried out 
 during AA's stint as a PDF at Physical Research Laboratory, Ahmedabad, 
 India.


\begin{thebibliography}{99}

\bibitem{Tulin:2017ara} 
  S.~Tulin and H.~B.~Yu,
  arXiv:1705.02358 [hep-ph].

\bibitem{Kleidis:2011ga} 
  K.~Kleidis and N.~K.~Spyrou,
  Astron.\ Astrophys.\  {\bf 529}, A26 (2011)
  doi:10.1051/0004-6361/201016057
  [arXiv:1104.0442 [gr-qc]].

\bibitem{Kleidis:2014xia} 
  K.~Kleidis and N.~K.~Spyrou,
  Astron.\ Astrophys.\  {\bf 576}, A23 (2015)
  doi:10.1051/0004-6361/201424402
  [arXiv:1411.6789 [astro-ph.CO]].

\bibitem{MNL2:MNL21082}
  A.~L.~Serra and  M.~J.~L.~D.~Romero,
Mon.\ Not.\ R.\ Astron.\ Soc.\ Lett.\ {\bf 415} , L74 (2011).
doi:10.1111/j.1745-3933.2011.01082.x

\bibitem{Calabrese:2009zza} 
  E.~Calabrese, M.~Migliaccio, L.~Pagano, G.~De Troia, A.~Melchiorri and P.~Natoli,
  Phys.\ Rev.\ D {\bf 80}, 063539 (2009).
  doi:10.1103/PhysRevD.80.063539

\bibitem{Murphy:1973zz} 
G.~L.~Murphy,
Phys.\ Rev.\ D {\bf 8}, 4231 (1973).

\bibitem{Padmanabhan:1987dg} 
T.~Padmanabhan and S.~M.~Chitre,

\bibitem{Fabris:2005ts} 
J.~C.~Fabris, S.~V.~B.~Goncalves and R.~de Sa Ribeiro,
Gen.\ Rel.\ Grav.\  {\bf 38}, 495 (2006)

\bibitem{Das:2008mj}
S.~Das and N.~Banerjee,
Int.\ J.\ Theor.\ Phys.\  {\bf 51}, 2771 (2012)

\bibitem{Gagnon:2011id} 
J.~S.~Gagnon and J.~Lesgourgues,
JCAP {\bf 1109}, 026 (2011)

\bibitem{Mathews:2008hk} 
G.~J.~Mathews, N.~Q.~Lan and C.~Kolda,
Phys.\ Rev.\ D {\bf 78}, 043525 (2008)

\bibitem{Li:2009mf} 
  B.~Li and J.~D.~Barrow,
  Phys.\ Rev.\ D {\bf 79}, 103521 (2009)
  doi:10.1103/PhysRevD.79.103521
  [arXiv:0902.3163 [gr-qc]].

\bibitem{Piattella:2011bs} 
  O.~F.~Piattella, J.~C.~Fabris and W.~Zimdahl,
  JCAP {\bf 1105}, 029 (2011)
  doi:10.1088/1475-7516/2011/05/029
  [arXiv:1103.1328 [astro-ph.CO]].

\bibitem{Velten:2011bg} 
  H.~Velten and D.~J.~Schwarz,
  JCAP {\bf 1109}, 016 (2011)
  doi:10.1088/1475-7516/2011/09/016
  [arXiv:1107.1143 [astro-ph.CO]].

\bibitem{Kaplinghat:2015aga} 
M.~Kaplinghat, S.~Tulin and H.~B.~Yu,
Phys.\ Rev.\ Lett.\  {\bf 116}, no. 4, 041302 (2016)  
  

\bibitem{Floerchinger:2014jsa}
S.~Floerchinger, N.~Tetradis and U.~A.~Wiedemann,
Phys.\ Rev.\ Lett.\  {\bf 114} (2015) no.9,  091301

\bibitem{Gavin:1985ph} 
S.~Gavin,
Nucl.\ Phys.\ A {\bf 435}, 826 (1985).

\bibitem{Kadam:2015xsa} 
G.~P.~Kadam and H.~Mishra,
Phys.\ Rev.\ C {\bf 92}, no. 3, 035203 (2015)


\bibitem{Ade:2013zuv} 
P.~A.~R.~Ade {\it et al.} [Planck Collaboration],
Astron.\ Astrophys.\  {\bf 571}, A16 (2014)

\bibitem{Weber:2009pt} 
M.~Weber and W.~de Boer,
Astron.\ Astrophys.\  {\bf 509}, A25 (2010)
doi:10.1051/0004-6361/200913381
[arXiv:0910.4272 [astro-ph.CO]].

\bibitem{KuziodeNaray:2007qi} 
R.~Kuzio de Naray, S.~S.~McGaugh and W.~J.~G.~de Blok,
Astrophys.\ J.\  {\bf 676}, 920 (2008)

\bibitem{Oh:2010ea} 
S.~H.~Oh, W.~J.~G.~de Blok, E.~Brinks, F.~Walter and R.~C.~Kennicutt, Jr,
Astron.\ J.\  {\bf 141}, 193 (2011)

\bibitem{Newman:2012nw} 
A.~B.~Newman, T.~Treu, R.~S.~Ellis and D.~J.~Sand,
Astrophys.\ J.\  {\bf 765}, 25 (2013)
\bibitem{Newman:2012nv} 
A.~B.~Newman, T.~Treu, R.~S.~Ellis, D.~J.~Sand, C.~Nipoti, J.~Richard and E.~Jullo,
Astrophys.\ J.\  {\bf 765}, 24 (2013)





\end{thebibliography}
\end{document}